\documentclass[9pt,a4paper,twoside]{tau-class/tau}
\usepackage[spanish]{babel}

\journalname{bor}
\title{Gravitational GateWave: Gravitational waves as LFO modulation amplitude}

\author[a]{Carlos Darío Badilla Cerdas }

\affil[a]{Escuela de Física, Universidad de Costa Rica}
\affil{\url{dari.badilla@gmail.com}}

\institution{Universidad de Costa Rica}
\footinfo{|borterion$\rangle$}

\institution{Universidad de Costa Rica}
\footinfo{borterion}

\begin{abstract}
We present a novel audio gate plugin that uses real gravitational‑wave strain data from the LIGO and Virgo detectors as low‑frequency oscillator (LFO) modulation sources. Conventional audio gates rely on synthetic waveforms—sine, triangle, sawtooth—whose fixed harmonic content produces predictable pumping effects. By contrast, gravitational‑wave chirps from compact binary coalescences sweep through frequency and amplitude on sub‑second timescales; when time‑scaled to LFO rates ($0.1$–$10$~Hz), these chirps yield modulation signals with continuously evolving harmonic structure. The plugin demonstrates that astrophysical signals can serve as musically compelling modulation sources, bridging sonification and creative audio processing.
\end{abstract}


\begin{document}
		
    \maketitle 
    \thispagestyle{firststyle} \tauabstract

\section{Introduction}
The audio gate is a fundamental dynamics processor that attenuates a signal when its level falls below a user‑defined threshold.  When the gate’s threshold is modulated by a low‑frequency oscillator (LFO), the familiar “pumping” or “breathing” effect heard in countless electronic music productions is obtained.  Conventionally, LFO waveforms are limited to standard geometric shapes – sine, triangle, sawtooth, square – whose harmonic content is static and musically predictable.  

As a physicist with idle hands and an interest in gravitational‑wave astronomy, I asked a simple question: what happens if the modulating waveform is not a mathematical abstraction but a real, high‑energy astrophysical signal?  The chirps recorded by the LIGO and Virgo detectors, produced by the inspiral and merger of black holes and neutron stars, sweep through frequency and amplitude in a matter of seconds.  When slowed down by a factor of a few hundred, these chirps become intricate, slowly evolving modulation sources whose characteristic “chirp” shape introduces a rich and continuously changing harmonic structure far beyond what synthetic LFOs can offer.

This paper describes a JUCE‑based audio plugin that implements such a concept: a gate processor whose LFO is fed by real gravitational‑wave strain data.  The plugin uses open‑access time series from the Gravitational‑Wave Open Science Center (GWOSC), processes them into loopable wavetables, and allows the user to select among several confirmed events.  I outline the signal‑processing chain, the software architecture, and the audible characteristics that make the gravitational‑wave gate both a creative tool and an educational demonstration of chirp signals.

\section{Background}
\subsection{Gravitational‑wave chirps}
Gravitational waves are ripples in spacetime emitted by accelerating masses, first directly detected in 2015 by the Advanced LIGO detectors \cite{abbott2016observation}.  The most luminous sources are compact binary coalescences – systems of two black holes, two neutron stars, or a mixed pair – whose inspiral, merger, and ringdown produce a characteristic “chirp” signal.  As the orbit shrinks, the orbital frequency and the emitted gravitational‑wave amplitude increase, creating a sweep that accelerates until the moment of merger.  The time‑domain waveform $h(t)$ for a non‑precessing, circular binary is well approximated by post‑Newtonian templates, exhibiting a monotonically rising frequency and amplitude envelope \cite{blanchet2014gravitational}.

The chirp’s defining features are its rapid frequency evolution and its broadband power.  For a typical stellar‑mass black‑hole binary like GW150914, the signal sweeps from $\approx 35\,\mathrm{Hz}$ to $\approx 250\,\mathrm{Hz}$ in about $0.2\,\mathrm{s}$.  When time‑scaled to LFO rates (below $20\,\mathrm{Hz}$), this sweep unfolds over several seconds, preserving the accelerating frequency profile that makes a chirp perceptually distinct from a simple linear glide.

\subsection{Low‑frequency oscillators in audio processing}
An LFO is an oscillator operating below the audio range (typically $0.01$–$20\,\mathrm{Hz}$) whose output modulates a parameter of an audio effect or synthesizer.  Standard waveforms – sine, triangle, pulse, sawtooth – produce periodic modulation with a fixed set of harmonics.  The spectral flatness and predictability of these shapes are precisely what makes them musically useful, yet they limit the complexity of the resulting timbral movement.  More advanced modulators employ wavetables, sample‑and‑hold, or chaotic maps, but the source material is almost always artificially generated.

Audio‑rate modulation of a gate has been explored to create aggressive distortion effects (often termed “trance gate” or “rhythmic gating”), but sub‑audio modulation remains dominated by simple periodic functions.  Using a real physical signal as an LFO introduces a deterministic but non‑synthetic evolution, combining a slow periodic loop with internal spectral motion.  Because gravitational‑wave chirps are naturally quasi‑periodic only when artificially looped, they offer a modulation timbre that sits between periodic and textural, opening new sound‑design possibilities.

\section{Methodology}
\subsection{Open data acquisition}
All gravitational‑wave strain data were obtained from the {GWOSC} The plugin currently includes 9 confirmed events from the first three observing runs (O1, O2, O3) of Advanced LIGO and Advanced Virgo {GWOSC}.  For each event, the 4‑kHz sampled strain time series of the Hanford (H1) and Livingston (L1) detectors were downloaded.  When both detectors had suitable data quality, the L1 stream was used, as it often exhibits a slightly higher signal‑to‑noise ratio for the selected events.

Each raw strain series was band‑passed between $20\,\mathrm{Hz}$ and $500\,\mathrm{Hz}$ using a 4th‑order Butterworth filter to remove seismic noise and high‑frequency artifacts.  The filtered signal was then centred by subtracting its mean.  Because the chirp is a transient, the analysis window was trimmed to a segment starting $0.1\,\mathrm{s}$ before the merger time (as reported in the GWTC catalogs) and ending $0.05\,\mathrm{s}$ after the merger, capturing the inspiral and the merger spike but excluding the low‑amplitude ringdown tail that would complicate looping.  The resulting snippets range from $0.15\,\mathrm{s}$ to $0.8\,\mathrm{s}$ in duration.

\subsection{Time scaling and normalization}
To transform a sub‑second chirp into a several‑second modulation cycle, a time‑stretching factor $S$ is applied.  The scaled time axis is $t' = S \cdot t$, with $S$ chosen so that the chirp duration $T_{\text{chirp}}$ maps to a user‑selectable LFO period $T_{\text{lfo}}$:
\[
S = \frac{T_{\text{lfo}}}{T_{\text{chirp}}}.
\]
For $T_{\text{lfo}} = 4\,\mathrm{s}$ and GW150914 ($T_{\text{chirp}} \approx 0.2\,\mathrm{s}$), $S = 20$.  Interpolation is performed using a cubic spline to avoid aliasing at the new effective sample rate.  After scaling, the waveform is normalised to the range $[-1, 1]$ so that the peak amplitude corresponds to the maximum modulation depth.  The root‑mean‑square (RMS) level is also computed and stored as an alternative normalisation target.

\subsection{Waveform representation}
The final preprocessed waveform is stored as a JUCE \texttt{AudioSampleBuffer} with a length of exactly $T_{\text{lfo}}$ at the plugin’s internal processing sample rate.  A small cross‑fade of $10\,\mathrm{ms}$ is applied at the loop boundary to suppress clicks when the waveform is read cyclically.  Because the chirp has a distinct start and end, looping introduces a discontinuity that, after cross‑fading, becomes an audible but musically acceptable “reset” transient, comparable to the loop point of a sampled instrument note.

\begin{figure} [H]
	\centering
	\includegraphics[width=0.9\linewidth]{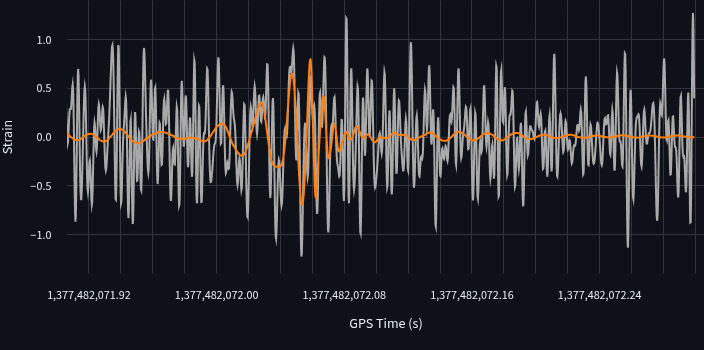}
	\caption{Waveform of GW230831 extracted, before and after filtering.}
	\label{fig:w}
\end{figure}

\section{Plugin Implementation}
\subsection{LFO engine}
The plugin is written in C++ using the JUCE \cite{juce} framework, targeting VST3 format.  The LFO engine is a \texttt{WavetableOscillator} that reads the stored gravitational‑wave buffer.  A phase accumulator increments by a step size determined by the desired LFO frequency $f_{\text{lfo}}$, with optional tempo‑sync to the host digital audio workstation (DAW).  The output of the LFO is a unipolar $[0,1]$ signal obtained by halving and offsetting the stored bipolar waveform.

\subsection{Parameter modulation}
The core audio path is a standard downward expander/gate with adjustable threshold, ratio, attack, hold, and release.  The LFO signal modulates the threshold \textbf{downward} from its static value.  If the static threshold is $T_0$ (in dB), the instantaneous threshold becomes
\[
T(t) = T_0 - M \cdot \text{LFO}(t) \cdot R,
\]
where $M$ is the modulation depth (0–100\%) and $R$ is a scaling factor mapping the unipolar LFO to a dB range (default $R = 30\,\mathrm{dB}$).  Thus, when the LFO reaches its maximum, the threshold is lowered by up to $30\,\mathrm{dB}$, opening the gate; when the LFO is at zero, the gate closes to the user‑set floor.  Attack, hold, and release times are kept fixed during modulation to preserve the envelope smoothing, although optional LFO‑modulation of the release time is available for additional rhythmic effects.

\subsection{Available gravitational‑wave events}
The plugin ships with an internal library of 8 events, selectable via a drop‑down menu:

\begin{itemize}
	\item GW150914 – the first detection, binary black hole.
	\item GW151226 – lighter binary black hole.
	\item GW170817 – binary neutron star merger (with a longer inspiral).
	\item GW190412 – asymmetric black‑hole binary.
	\item GW190521 – the most massive black‑hole merger observed to date.
	\item GW191103 – \textit{Star Cycle}.
	\item GW230831 – \textit{Heredia, august 31st}. 
	\item GW231126 – \textit{Whole new world}.
\end{itemize}

\section{Results}

\subsection{Example waveforms}
The Figure \ref{fig:w} shows that the gravitational‑wave chirp exhibits a long, gradual rise that accelerates into a sharp peak, followed by a rapid fall.  The power spectrum of the chirp waveform is broadband, with a formant‑like emphasis around the frequency where the chirp spends most of its time.  In contrast, the sine and triangle have line spectra, illustrating why the chirp provides a richer harmonic modulation.

\subsection{Audible characteristics}
When applied to sustained synth pads, drum loops, or full mixes, the gravitational‑wave gate produces a “pumping” effect that evolves timbrally over the cycle.  Three perceptual features stand out:

\begin{enumerate}
	\item \textbf{Dynamic harmonic emphasis.}  Because the chirp’s instantaneous frequency changes throughout the cycle, the rate at which the gate opens and closes is not constant.  Fast‑changing sections produce a transient, almost granular texture, while slow sections create a smooth swelling.
	\item \textbf{Event‑dependent character.}  Neutron‑star chirps (e.g., GW170817) sweep over a longer duration and lower frequency range, resulting in a gentler, more sinusoidal modulation.  The massive GW190521 chirp is extremely brief even after scaling, yielding a percussive, staccato modulation.  Users can select an event to match the rhythmic feel of the material.
	\item \textbf{Harmonic sidebands.}  The asymmetric shape of the chirp introduces even and odd harmonics into the modulation signal, which translate into sidebands around the carrier frequencies present in the audio.  When the gate’s threshold is pushed into extreme settings, the sidebands become clearly audible as a form of ring‑modulation‑like coloration.
\end{enumerate}

\section{Discussion}
\subsection{Educational applications}
The plugin serves as an intuitive tool for explaining chirp signals.  Users who would not otherwise encounter gravitational‑wave data can see the waveform, hear its influence on sound, and connect the audible sweep to the astrophysical inspiral.  A dedicated “educational” panel overlays the waveform with the inspiral, merger, and ringdown phases labeled, and provides links to the original GWOSC event pages.  This bridges the gap between abstract physics and tangible musical experience, making the plugin suitable for outreach workshops and physics demonstrations.

\subsection{Limitations}
Several limitations should be noted.  First, the gravitational‑wave waveforms are short transients; when looped, the repetitive “reset” click, although cross‑faded, may become obtrusive at low LFO rates.  Future versions could use a phase‑vocoder to smoothly re‑synthesize the waveform without a hard loop point.  Second, the plugin does not perform real‑time gravitational‑wave sonification; it relies on pre‑processed event buffers and thus cannot respond to new detections without a software update.  Third, the time‑scaling is linear, which preserves the chirp’s frequency‑time profile but may not provide the same perceptual smoothness at all LFO rates; non‑linear warping might offer more musical control.  

\section{Conclusion}
A gate audio effect whose LFO is driven by real gravitational‑wave chirps has been implemented as a cross‑platform JUCE plugin.  By time‑scaling open data from LIGO and Virgo, the chirp waveforms are transformed into slowly evolving modulation sources that impart a distinct harmonic character to the familiar pumping effect.  The plugin makes gravitational‑wave signals accessible to musicians and educators alike, demonstrating that even the most esoteric physical phenomena can find a creative home in the studio.  Ongoing work includes adding more events from the GWTC‑3 catalog, supporting real‑time event alerts, and extending the approach to other dynamics processors such as compressors and tremolos.

\begin{quote}
	The gravity of memories \\
	drags me into the sea waves. \\
	Perhaps by staying, I sink deeper. \\
	My face, distorted like a conformal map.
\end{quote}

\section*{Acknowledgements}
The author thanks the LIGO Scientific Collaboration and the Virgo Collaboration for making gravitational‑wave data publicly available, and the JUCE team for a robust audio framework.  This work made use of the Gravitational‑Wave Open Science Center (gw‑openscience.org), a service of LIGO Laboratory, the LIGO Scientific Collaboration, and the Virgo Collaboration.  


\addcontentsline{none}{section}{References}
\printbibliography


\end{document}